\documentstyle[prd,aps,floats,twocolumn]{revtex}

\def\H{{\cal H}}
\def\be{\begin{equation}}
\def\ee{\end{equation}}
\def\bea{\begin{eqnarray}}
\def\eea{\end{eqnarray}}

\newcommand\lsim{\mathrel{\rlap{\lower4pt\hbox{\hskip1pt$\sim$}}
    \raise1pt\hbox{$<$}}}
\newcommand\gsim{\mathrel{\rlap{\lower4pt\hbox{\hskip1pt$\sim$}}
    \raise1pt\hbox{$>$}}}

\def\calp{{\cal P}}

\newcommand\bfx{{\bf x}}

\begin{document}
\preprint{} 
\draft

%
%
\input epsf
\renewcommand{\topfraction}{0.99}
\renewcommand{\bottomfraction}{0.99}
\twocolumn[\hsize\textwidth\columnwidth\hsize\csname@twocolumnfalse\endcsname

\title{On Non-Gaussianity in the Curvaton Scenario}
\author{N. Bartolo$^1$, S. Matarrese$^2$ and A. Riotto$^2$}
\address{(1)Astronomy Centre, University of Sussex
Falmer, Brighton, BN1 9QJ, U.K.}
\address{(2) Department of Physics and INFN
Sezione di Padova, via Marzolo 8, I-35131 Padova, Italy}

\date{\today}
\maketitle
\begin{abstract}
\noindent
Since a positive future detection of non-linearity
in the cosmic microwave  background anisotropy pattern might allow
to descriminate among different  mechanisms
giving rise to   cosmological adiabatic perturbations,
we study the evolution of the second-order 
cosmological  curvature perturbation on super-horizon scales
in the curvaton scenario. We provide
the exact expression for the  
non-Gaussianity in  the 
primordial perturbations
including gravitational 
second-order corrections which are particularly relevant
in the case in which
the curvaton dominates the energy density before it decays.
As a byproduct, we show that in the standard
scenario where cosmological curvature perturbations are induced by
the inflaton field, 
the second-order curvature perturbation is conserved 
even during the reheating stage after inflation.
\end{abstract}

\pacs{PACS numbers: 98.80.cq; DFPD-A-03-33}

\vskip2pc]

\section{Introduction}

One of the basic ideas of modern cosmology is that there was an epoch early
in the history of the universe when potential, or vacuum, energy 
associated to a scalar field, the inflaton, 
dominated other forms of energy density such as matter or radiation. 
During such a
vacuum-dominated era the scale factor grew exponentially (or nearly
exponentially) in time. During this phase, dubbed inflation 
\cite{guth81,lrreview},
a small,  smooth spatial region of size less than the Hubble radius
could grow so large as to easily encompass the comoving volume of the 
entire presently observable universe. If the universe underwent
such a period of rapid expansion, one can understand why the observed
universe is so homogeneous and isotropic to high accuracy.

Inflation has also become the dominant 
paradigm for understanding the 
initial conditions for structure formation and for Cosmic
Microwave Background (CMB) anisotropy. In the
inflationary picture, primordial density and gravity-wave fluctuations are
created from quantum fluctuations ``redshifted'' out of the horizon during an
early period of superluminal expansion of the universe, where they
are ``frozen'' \cite{muk81,hawking82,starobinsky82,guth82,bardeen83}. 
Perturbations at the surface of last scattering are observable as temperature 
anisotropy in the CMB, which was first detected by the Cosmic Background 
Explorer (COBE) satellite \cite{smoot92,bennett96,gorski96}.
The last and most impressive confirmation of the inflationary paradigm has 
been recently provided by the data 
of the Wilkinson Microwave Anistropy Probe (WMAP) mission which has 
marked the beginning of the precision era of the CMB measurements in space
\cite{wmap1}.
The WMAP collaboration has  produced a full-sky map of the angular variations 
of the CMB, with unprecedented accuracy.
WMAP data confirm the inflationary mechanism as responsible for the
generation of curvature (adiabatic) super-horizon fluctuations. 

Despite the simplicity of the inflationary paradigm, the mechanism
by which  cosmological adiabatic perturbations are generated  is not
yet fully established. In the standard picture, the observed density 
perturbations are due to fluctuations of the inflaton field itself. 
When inflation ends, the inflaton oscillates about the minimum of its
potential and decays, thereby reheating the universe. As a result of the 
fluctuations
each region of the universe goes through the same history but at slightly
different times. The 
final temperature anisotropies are caused by the fact that
inflation lasts different amounts of time in different regions of the universe
leading to adiabatic perturbations. Under this hypothesis, 
the WMAP dataset already allows
to extract the parameters relevant 
for distinguishing among single-field inflation models \cite{ex}.

An alternative to the standard scenario is represented by the curvaton 
mechanism
\cite{curvaton1,LW,curvaton3,LUW} where the final curvature perturbations
are produced from an initial isocurvature perturbation associated to the
quantum fluctuations of a light scalar field (other than the inflaton), 
the curvaton, whose energy density is negligible during inflation. The 
curvaton isocurvature perturbations are transformed into adiabatic
ones when the curvaton decays into radiation much after the end 
of inflation\footnote{
Recently, another mechanism for the generation of cosmological
perturbations has been proposed \cite{gamma}.  
It acts during the reheating
stage after inflation if super-horizon spatial
fluctuations in the decay rate of the inflaton field
are induced during inflation, causing  adiabatic perturbations
in  the final reheating temperature
in different regions of the universe.}.
Contrary to the standard picture, the curvaton 
mechanism exploits the fact that 
the total curvature perturbation (on uniform density hypersurfaces)
$\zeta$ can change on arbitrarily large scales due to a non-adiabatic
pressure perturbation    which may be 
present  in a multi-fluid system \cite{Mollerach,MFB,GBW,WMLL,wmu}.
While the entropy
perturbations evolve independently of the curvature perturbation on
large scales,  the evolution of the large-scale curvature is
sourced by entropy perturbations. 

Fortunately, the standard and the curvaton  scenarios 
have different observational signatures. 
The curvaton scenario 
allows to generate the observed level of density perturbations with a 
much lower scale of inflation and thus generically predicts 
a smaller level of gravitational waves. 
More interestingly,  density perturbations generated through the
curvaton scenario 
could be highly non-Gaussian and  the   level of non-Gaussianity
in the
primordial perturbations,  which is usually parametrized by
a dimensionless non-linear parameter $f_{\rm NL}$, 
depends upon
an unknown parameter $r$ indicating
the fraction of energy density contributed by the curvaton field
at the epoch of its decay. For tiny values of $r$, it has been 
estimated that  
the non-Gaussianity can be large enough to be 
detectable by  present  CMB experiments \cite{LW,LUW}; 
the current WMAP \cite{k} bound on
non-Gaussianity, $\vert f_{\rm NL} \vert \lsim 10^2$, 
already requires $r$ to be larger than about $10^{-2}$.

This curvaton prediction has to be contrasted to what 
predicted within the traditional one-single field model
of inflation where
the initially tiny non-linearity in the cosmological  perturbations
generated during the inflationary epoch
\cite{noi,maldacena}
gets enhanced in the post-inflationary stages giving rise to
a well-defined prediction for the non-linearity in the gravitational 
potentials~\cite{BMR}. 

Since a positive future detection of non-linearity
in the CMB anisotropy pattern might allow
to descriminate among the mechanisms
by which  cosmological adiabatic perturbations are generated, it is
clear that the precise determination of the non-Gaussianity 
predicted by the curvaton mechanism is of primary interest. 

The goal of this paper is to provide an exact expression for the 
non-linear parameter $f_{\rm NL}$ within the curvaton scenario
including second-order corrections from gravity 
which are particularly relevant
in the case in which
the curvaton dominates the energy density before it decays. 
We perform a fully relativistic analysis of the dynamics
of second-order perturbations taking advantage of 
the second-order gauge-invariant curvature 
perturbation introduced in Refs. \cite{lw,mw} (see also
\cite{noi,nak,rigo}) and showing how it evolves
on arbitrarily large scales in the presence of two fluids, matter (the
curvaton) and radiation. 
Our results generalize the estimates given in Refs. \cite{LW,LUW}
and confirm their findings in the limit $r\ll 1$.

The paper is organized as follows. In Section II we briefly summarize the
properties of the curvaton scenario and how the primordial curvature
perturbations are created at first-order. In section III we compute the 
second-order curvature perturbation from the curvaton fluctuations
and determine the exact expression for the non-linear parameter $f_{\rm NL}$
as a function of the unknown parameter $r$. In section IV we 
show that our findings can be easily generalized to the standard
scenario where adiabatic perturbations are
provided by the same field driving inflation and 
prove that the second-order curvature perturbation is conserved 
even during the reheating stage after inflation when the inflaton field
decays to give birth to the standard radiation phase. Finally, Section V 
contains our conclusions.

\section{Generating the curvature perturbation at linear order}
During inflation the curvaton field $\sigma$ is supposed to give a 
negligible contribution to the energy density and to be an almost free 
scalar field,  with a small effective mass 
$m^2_\sigma=|\partial^2 V/\partial \sigma^2| \ll H_I^2$ \cite{LW,LUW}, where
$H_I=\dot{a}/a$ is the Hubble rate during inflation, $a$ is the scale factor
and a dot denotes derivative with respect to cosmic time.

The unperturbed curvaton field satisfies the equation of motion
\be
\label{backg}
\sigma''+2 \H \sigma'+a^2 \frac{\partial V}{\partial \sigma}=0\, ,
\ee
where a prime denotes differentation with respect to the conformal time 
$d\tau=dt/a$ and $\H=a'/a$ is the Hubble parameter in conformal time.
It is also usually assumed that the curvaton field is very weakly coupled to 
the scalar fields driving inflation and that the curvature perturbation 
from the inflaton fluctuations is negligible
\cite{LW,LUW}. Thus, if we expand the curvaton field  up to first-order 
in the 
perturbations around the homogeneous background as $\sigma(\tau,\bfx)
=\sigma(\tau)+\delta^{(1)}\sigma$, the 
linear perturbations satisfy on large scales
\be
\label{sigma1}
\delta^{(1)} \sigma ''+2 \H \delta^{(1)} \sigma' +a^2 
\frac{\partial^2 V}{\partial \sigma^2}\, \delta^{(1)} \sigma=0\, .
\ee
As a result on superhorizon scales its fluctuations 
$\delta \sigma$ will be Gaussian distributed and with a nearly 
scale-invariant spectrum given by 
\be
\calp_{\delta\sigma}^\frac12(k) \approx \frac{H_*}{2\pi}
\label{pinf}
\,,
\ee
where the subscript $*$ denotes the epoch of horizon exit $k=aH$.     
Once inflation is over the inflaton energy density will be converted to 
radiation ($\gamma$) and the curvaton field will remain approximately 
constant until $H^2 \sim m_\sigma^2$. At this epoch the curvaton field begins 
to oscillate around the minimum of its potential which can be safely 
approximated to be quadratic $V \approx \frac{1}{2} m_\sigma^2 \sigma^2$.
During this stage the energy density of the curvaton field just scales as 
non-relativistic matter $\rho_\sigma \propto a^{-3}$. 
The energy density in the oscillating field is
\be
\label{energyoscill}
\rho_\sigma(\tau,\bfx) \approx
m_\sigma^2 \sigma^2(\tau,\bfx)
\,,
\label{rhosigma}
\ee
and it can be expanded into a homogeneous background 
$\rho_\sigma(\tau)$ and a first-order perturbation $\delta^{(1)} 
\rho_\sigma$ as 
\be
\label{rhocurv1}
\rho_\sigma(\tau,\bfx)=\rho_\sigma(\tau)+\delta^{(1)} \rho_\sigma(\tau,\bfx)=
m_\sigma^2 \sigma+2 m_\sigma^2\, \sigma \, \delta^{(1)} \sigma
\, . 
\ee
As it follows from Eqs.~(\ref{backg}) and (\ref{sigma1}) for a 
quadratic potential the ratio $\delta^{(1)} \sigma/\sigma$ remains 
constant and     
the resulting relative energy density perturbation is
\be
\label{relrhocurv}
\frac{\delta^{(1)}\rho_\sigma}{\rho_\sigma}=2 \left(
\frac{\delta^{(1)} \sigma}{\sigma} \right)_* \, ,
\ee
where the $*$ stands for the value at  horizon crossing.

Such perturbations in the energy density of the 
curvaton field produce in fact a primordial density 
perturbation well after the end of inflation.     
The primordial adiabatic density perturbation is associated with a 
perturbation in the spatial curvature $\psi$ and it is usually characterized 
in a gauge-invariant manner by the curvature perturbation $\zeta$ on 
hypersurfaces of uniform total density $\rho$. At linear order the quantity 
$\zeta$ is given by the gauge-invariant formula \cite{Bardeen}
\be
\label{zetatot}
\zeta^{(1)}= -\psi^{(1)} - \H\frac{\delta^{(1)}\rho}{\rho'}
\, ,
\label{eqzeta}
\ee     
and on large scales it obeys the equation of motion \cite{Bardeen,WMLL} 
\begin{equation}
 \label{zetadot}
 \zeta^{(1)'} = - { \H \over \rho +P} \,\delta^{(1)} P_{\rm nad} \,,
\end{equation}
where $\delta^{(1)} P_{\rm nad}=\delta^{(1)} P - c_s^2\delta^{(1)}\rho$ is 
the non-adiabatic pressure perturbation, $\delta^{(1)} P$ being the pressure 
perturbation and $c_s^2=P'/\rho'$  the adiabatic sound speed. 
In the curvaton scenario the curvature perturbation is generated well after 
the end of inflation during the oscillations of the curvaton field because 
the pressure of the mixture of matter (curvaton) and radiation produced by 
the inflaton decay is not adiabatic. A convenient way to study this 
mechanism is 
to consider the curvature perturbations $\zeta_i$ associated with each 
individual energy density components, which to linear order are defined as
\cite{WMLL}

\begin{equation}
\label{zetai}
\zeta^{(1)}_{i}
 \equiv - \psi^{(1)} - \H\left(\frac{\delta^{(1)}\rho_{i}}
{\rho_{i}'}\right)
\,.
\end{equation}   
Therefore, during the oscillations of the curvaton field,  
the total curvature perturbation in 
Eq.~(\ref{zetatot}) can be written as a weighted sum of the single 
curvature perturbations \cite{WMLL,LUW}
\be
\label{zetasum}
\zeta^{(1)}=(1-f)\zeta^{(1)}_\gamma+f\zeta^{(1)}_\sigma \, ,
\ee
where the quantity 
\begin{equation}
\label{deff}
 f = \frac{3\rho_\sigma}{4\rho_\gamma +3\rho_\sigma} 
\end{equation}
defines the relative contribution of the curvaton field 
to the total curvature 
perturbation. From now on we shall work under the approximation of  
sudden decay of the curvaton field. Under this approximation the curvaton and 
the radiation components $\rho_\sigma$ and $\rho_\gamma$ satisfy 
separately the energy conservation equations 
\bea
\label{conseqs}
\rho_\gamma'=-4 \H \rho_\gamma\, ,\nonumber \\
\rho_\sigma'=-3 \H \rho_\sigma \, ,
\eea
and the curvature 
perturbations $\zeta_i$ remains constant on superhorizon scales until 
the decay of the curvaton. Therefore from Eq.~(\ref{zetasum}) it follows 
that the first-oder curvature pertubation evolves on large scales as
\be
\zeta^{(1)'}=f'(\zeta^{(1)}_\sigma-\zeta^{(1)}_\gamma)
=\H f(1-f)(\zeta^{(1)}_\sigma-\zeta^{(1)}_\gamma)\, ,
\ee 
and by comparison with Eq.~(\ref{zetadot}) one obtains the expression for the
non-adiabatic pressure perturbation at first order \cite{LW,LUW}
\be
\label{pressurepert}
\delta^{(1)} P_{\rm nad}=\rho_\sigma (1-f)(\zeta^{(1)}_\gamma-
\zeta^{(1)}_\sigma)\, .
\ee 
Since in the curvaton scenario 
it is supposed that the curvature perturbation 
in the radiation produced at the end of inflation is negligible 
\be
\label{zetagamma}
\zeta^{(1)}_\gamma=-\psi^{(1)}+\frac{1}{4}\frac{\delta^{(1)} \rho_\gamma}
{\rho_\gamma}=0 \, .
\ee
Similarly the value of $\zeta^{(1)}_\sigma$ is fixed by 
the fluctuations of  the curvaton during inflation
\be
\label{zetasigma}
\zeta^{(1)}_\sigma=-\psi^{(1)}+\frac{1}{3}
\frac{\delta^{(1)}\rho_\sigma}{\rho_\sigma}= 
\zeta^{(1)}_{\sigma I}\, ,
\ee 
where $I$ stands for the value of the 
fluctuations during inflation.  
From Eq.~(\ref{zetasum}) the total curvature perturbation 
during the curvaton oscillations is given by  
\be
\label{zetaoscill}
\zeta^{(1)}=f \zeta^{(1)}_\sigma \, . 
\ee
As it is clear from Eq.~(\ref{zetaoscill}) initially, 
when the curvaton energy density is subdominant, the 
density perturbation in the curvaton field $\zeta^{(1)}_\sigma$ gives a 
negligible contribution to the total curvature perturbation, 
thus corresponding to an isocurvature (or entropy) perturbation. 
On the other hand during the oscillations $\rho_\sigma \propto a^{-3}$ 
increases with respect to the energy density of radiation 
$\rho_\gamma\propto a^{-4}$, and the perturbations in the curvaton field 
are then converted into the curvature perturbation.     
Well after the decay of the curvaton, 
during the conventional radiation and 
matter dominated eras, the total curvature perturbation  
will remain constant on superhorizon scales at a value which, 
in the sudden decay approximation, is fixed by Eq.~(\ref{zetaoscill}) at 
the epoch of curvaton decay
\be
\label{atcurvdecay}
\zeta^{(1)}=f_D\, \zeta^{(1)}_\sigma \, ,
\ee
where $D$ stands for the epoch of the curvaton decay.

Going beyond the sudden decay approximation it is possible to introduce a 
transfer parameter $r$ defined as \cite{LUW,wmu}
\be
\zeta^{(1)}=r\zeta^{(1)}_\sigma \, ,
\ee  
where $\zeta^{(1)}$ is evaluated well after the epoch of the curvaton 
decay and $\zeta^{(1)}_\sigma$ is evaluated well before this epoch.
The numerical study of the coupled perturbation equations has been performed 
in Ref.~\cite{wmu} showing that the sudden decay approximation is exact when 
the curvaton dominates the energy density before it decays $(r=1)$, while 
in the opposite case   
\be
r\approx \left( \frac{\rho_\sigma}{\rho} \right)_D.
\ee


\section{Second-order curvature perturbation from the curvaton 
fluctuations}
Here we generalize to second-order in the density perturbations
the results of the previous section.

As it has been shown in Ref.~\cite{mw} it is possible to define the 
second-order curvature  perturbation on uniform total density 
hypersurfaces by the quantity  (up to a gradient term) 
\bea
\label{qqq}
\zeta^{(2)}&=&-\psi^{(2)}-\H\frac{\delta^{(2)}\rho}{\rho^\prime}\nonumber \\
&+&2\H\frac{\delta^{(1)}\rho^\prime}{\rho^\prime}
\frac{\delta^{(1)}\rho}{\rho^\prime}
+2\frac{\delta^{(1)}\rho}{\rho^\prime}\left(\psi^{(1)\prime}
+ 2\H\psi^{(1)}\right) \nonumber \\
&-&\left(\frac{\delta^{(1)}\rho}{\rho^\prime}
\right)^2 \left(\H \frac{\rho^{\prime\prime}}{\rho^\prime}-
\H^\prime-2\H^2\right) \, ,
\eea
where the curvature perturbation $\psi$ has been expanded up to 
second-order as $\psi=\psi^{(1)}+\frac{1}{2} \psi^{(2)}$ and 
$\delta^{(2)} \rho$ 
corresponds to the second-order perturbation in the total 
energy density around the homogeneous background  $\rho(\tau)$ 
\bea
\label{rho}
\rho(\tau, \bfx)&=&\rho(\tau)+\delta\rho(\tau, \bfx)\nonumber \\
&=&\rho(\tau)+
\delta^{(1)}\rho(\tau, \bfx)+\frac{1}{2}\delta^{(2)}
\rho(\tau, \bfx)\, .
\eea
The quantity $\zeta^{(2)}$ is gauge-invariant and, as its first-order 
counterpart defined in Eq.~(\ref{zetatot}),  
it is sourced on superhorizon scales 
by a second-order non-adiabatic pressure perturbation \cite{mw}. 

In Ref.~\cite{BMR} 
the conserved quantity $\zeta^{(2)}$ has been used in the standard scenario
where the generation of cosmological perturbations is induced by 
fluctuations of the inflaton field (and there is no curvaton)
in order to follow the evolution  
on large scales of 
the primordial non-linearity in the cosmological perturbations from a period 
inflation to the matter dominate era.   
In the present scenario the conversion of the 
curvaton isocurvature perturbations into a final curvature perturbation 
at the epoch of the curvaton decay can be followed through the sum 
(\ref{zetasum}) of the 
individual curvature perturbations weighted by the ratio $f$ of 
Eq.(\ref{deff}).  

Let us now extend such a result at second order in the perturbations.
As we shall see in Sec.~\ref{non-gauss} this result will enable us to 
compute in an exact way the level of non-Gaussianity produced by the 
non-linearity of the perturbations in the curvaton energy density.

Since the quantities $\zeta^{(1)}_i$ and $\zeta^{(2)}_i$  
are gauge-invariant, we choose to 
work in the spatially flat gauge $\psi=0$ if not otherwise specified.
Note that from Eqs.~(\ref{relrhocurv}) and (\ref{zetasigma})
the value of $\zeta^{(1)}_\sigma$ is thus given by
\be
\label{zetasigmainfl}
\zeta^{(1)}_\sigma=\frac{1}{3} \frac{\delta^{(1)}\rho_\sigma}{\rho_\sigma}=
\frac{2}{3} \frac{\delta^{(1)} \sigma}{\sigma}=
\frac{2}{3} \left( \frac{\delta^{(1)} \sigma}{\sigma} \right)_* \, ,
\ee 
where we have used the fact that $\zeta^{(1)}_\sigma$ (
or equivalently $\delta^{(1)} \sigma/ \sigma$) remains constant, while
from Eq.~(\ref{zetasigma}) in the spatially flat gauge
\be
\label{zetaroscill}
\zeta^{(1)}_\gamma=\frac{1}{4}
\frac{\delta^{(1)} \rho_\gamma}{\rho_\gamma} \, .
\ee
During the oscillations of the curvaton field
the first-order energy conservation equations 
in the spatially flat gauge $\psi=0$ yield on large scales\footnote{
Here and in  the following
we neglect  gradient terms  which, upon integration over time, may
give rise to non-local operators which are not necessarily 
suppressed on large-scales
being of the form  $\nabla^{-2}\left[\nabla(\cdot)\nabla(\cdot)\right]$ or
$\nabla^{-2}\left[(\cdot)\nabla^2(\cdot)\right]$. In this paper
we will focus on the momentum-independent part of the non-linear
parameter $f_{\rm NL}$.}
 
\be
\label{constot}
\delta^{(1)} \rho'=\delta^{(1)} \rho_\sigma'+\delta^{(1)} \rho_\gamma'=
-3 \H\delta^{(1)} \rho_\sigma-4 \H \delta^{(1)} \rho_\gamma\, ,
\ee
and hence using Eqs.~(\ref{conseqs}), (\ref{zetasigmainfl}), 
(\ref{zetaroscill}) and 
(\ref{constot}) 
\bea
\frac{\delta^{(1)} \rho'}{\rho'}&=&3f\zeta^{(1)}_\sigma+4(1-f) 
\zeta^{(1)}_\gamma\, ,\nonumber \\
\H \frac{\delta^{(1)} \rho}{\rho'}&=&- f \zeta^{(1)}_\sigma- (1-f) 
\zeta^{(1)}_\gamma\, .
\eea
We can thus rewrite the total second-order 
curvature perturbation $\zeta^{(2)}$ as
\bea
\zeta^{(2)}&=&-\H \frac{\delta^{(2)} \rho}{\rho'} \nonumber \\
&-&\left[ f \zeta^{(1)}_\sigma+(1-f) \zeta^{(1)}_\gamma\right] 
\left[ f^2 \zeta^{(1)}_\sigma+(1-f)(2+f) \zeta^{(1)}_\gamma\right] 
\, . \nonumber
\\
&&
\eea
In a similar manner to the linear order, let us introduce now the 
curvature perturbations $\zeta^{(2)}_i$ at second order for each 
individual component. Such quantities will be given by the same formula 
as Eq.~(\ref{qqq}) relatively to each energy density $\rho_i$ 
\bea
\label{zeta2singole}
\zeta^{(2)}_i&=&-\psi^{(2)}-\H\frac{\delta^{(2)}\rho_i}
{\rho_{i}^\prime}\nonumber \\
&+&2\H\frac{\delta^{(1)}\rho_{i}^\prime}{\rho_{i}^\prime}
\frac{\delta^{(1)}\rho_i}{\rho_{i}^\prime}
+2\frac{\delta^{(1)}\rho_i}{\rho_{i}^\prime}\left(\psi^{(1)\prime}
+ 2\H\psi^{(1)}\right) \nonumber \\
&-&\left(\frac{\delta^{(1)}\rho_i}{\rho_{i}^\prime}
\right)^2 \left(\H \frac{\rho_{i}^{\prime\prime}}{\rho_{i}^\prime}-
\H^\prime-2\H^2\right) \, .
\eea

Using the 
same procedure described above it follows that in the spatially flat gauge 
\bea
\label{zetass}
\zeta^{(2)}_\sigma&=&\frac{1}{3}\frac{\delta^{(2)} \rho_\sigma}{\rho_\sigma}-
\left( \zeta^{(1)}_\sigma \right)^2 \, , \\ 
\label{zetagammass}
\zeta^{(2)}_\gamma&=&\frac{1}{4} \frac{\delta^{(2)} 
\rho_\gamma}{\rho_\gamma}- 2
\left( \zeta^{(1)}_\gamma\right)^2\, .
\eea    
Such quantities are gauge-invariant and, in the sudden decay approximation 
they are separately conserved until the curvaton decay. 
Using Eqs.~(\ref{zetass}) and (\ref{zetagammass}) to express 
the second-order perturbation 
in the total energy density $ \delta^{(2)} \rho=\delta^{(2)}\rho_\sigma+
\delta^{(2)}\rho_\gamma$, and after some algebra, one finally obtains 
the following expression for the total curvature perturbation 
$\zeta^{(2)}$
\bea
\label{main}
\zeta^{(2)}&=&f \zeta^{(2)}_\sigma+(1-f) \zeta^{(2)}_\gamma\nonumber \\
&+&f(1-f)(1+f)\left( \zeta^{(1)}_\sigma-\zeta^{(1)}_\gamma\right)^2\, .
\eea
Eq.~(\ref{main}) is one of our main results. It generalizes to second-order 
in the perturbations the weighted sum of Eq.~(\ref{zetasum}). In particular 
notice that in the limit where one of the two fluid is completely subdominant
($f \rightarrow 0$ or $f \rightarrow 1$) the corresponding curvature 
perturbation $\zeta^{(2)}_i$ turns out to coincide with the total one 
$\zeta^{(2)}$. 

Under the sudden decay approximation of the curvaton field the individual 
curvature perturbations are separately conserved on large scales, 
and thus from Eq.~(\ref{main}) it follows that $\zeta^{(2)}$ evolves 
according to the equation
\be
\label{evolzeta2}
\zeta^{(2)'}=f'\left( \zeta^{(2)}_\sigma-\zeta^{(2)}_\gamma\right)
+f' (1-3f^2) \left( \zeta^{(1)}_\sigma-\zeta^{(1)}_\gamma\right)^2\, .
\ee
Note that Eq.~(\ref{evolzeta2}) can be rewritten as ~\cite{mw}  
\bea
\zeta^{(2)'}&=&- \frac{\H}{\rho+P} \widetilde{\delta^{(2)}}P \nonumber \\
&-&\frac{2}{\rho+P}
\left[ \delta^{(1)}P_{\rm nad} -2(\rho+P) \zeta^{(1)} \right] 
\zeta^{(1)'} \, ,
\eea 
with  $\delta^{(1)} P_{\rm nad}$ given by Eq.~(\ref{pressurepert}) and
\bea
\widetilde{\delta^{(2)}}P&=&\rho_{\sigma}(1-f)
\Big[ \left( \zeta^{(2)}_\gamma-\zeta^{(2)}_\sigma \right) 
+ (f^2+6f-1)\nonumber\\
&\times&\left( \zeta^{(1)}_\sigma-\zeta^{(1)}_\gamma \right )^2 
+ 4\zeta^{(1)}_\gamma \left( \zeta^{(1)}_\sigma-\zeta^{(1)}_\gamma
\right) \Big]\, ,
\eea
is the gauge-invariant non-adiabatic pressure perturbation 
on uniform density hypersurfaces on large scales which
can be checked to coincide with the generic expression provided
in Ref. ~\cite{mw}

\begin{eqnarray}
\widetilde{\delta^{(2)}}P&=&\delta^{(2)}P-\frac{P'}{\rho'}\delta^{(2)}\rho
+P'\left[2\left(\frac{\delta^{(1)'}\rho}{\rho'}-
\frac{\delta^{(1)'}P}{P'}\right)\frac{\delta^{(1)}\rho}{\rho'}\right.
\nonumber\\
&+&\left.\left(\frac{P''}{P}-\frac{\rho''}{\rho}\right)
\left(\frac{\delta^{(1)}\rho}{\rho'}\right)^2\right]\, .
\end{eqnarray}
The second-order curvature perturbation in the standard 
radiation or matter eras will remain constant on superhorizon scales and, 
in the sudden decay approximation, it is thus given by 
the quantity in Eq.~(\ref{main}) evaluated at the epoch of the curvaton 
decay
\be
\label{zeta2decay}
\zeta^{(2)}=f_D \zeta^{(2)}_\sigma+f_D \left( 1-f^2_D \right) \left( 
\zeta^{(1)}_\sigma \right)^2 \, ,
\ee 
where we have used the curvaton hypothesis that the curvature perturbation
in the radiation produced at the end of inflation is negligible so that 
$\zeta^{(1)}_\gamma\approx 0$ and 
$\zeta^{(2)}_\gamma\approx 0$.
The curvature perturbation $\zeta^{(1)}_\sigma$ is given 
by Eq.~(\ref{zetasigmainfl}), while
$\zeta^{(2)}_\sigma$ in Eq.~(\ref{zetass}) is obtained by   
expanding the energy density of the curvaton field, Eq.~(\ref{energyoscill}),
up to second order in the curvaton fluctuations
\bea
\rho_\sigma(\bfx,t)&=&\rho_\sigma(\tau)+
\delta^{(1)}\rho_\sigma(\tau, x^i)+\frac{1}{2}\delta^{(2)}
\rho_\sigma(\tau, x^i)\nonumber \\
&=&m_\sigma^2 \sigma+
2 m_\sigma^2\, \sigma \, \delta^{(1)} \sigma+m^2_\sigma \left( 
\delta^{(1)} \sigma \right)^2\, .
\eea
It follows that
\be
\frac{\delta^{(2)} \rho_\sigma}{\rho_\sigma}=\frac{1}{2} 
\left( \frac{\delta^{(1)}\rho_\sigma}{\rho_\sigma} \right)^2
=\frac{9}{2} \left( \zeta^{(1)}_\sigma \right)^2\, ,
\ee
where we have used Eq.~(\ref{zetasigmainfl}), and hence from 
Eq.~(\ref{zetass}) we obtain
\be
\label{zeta2infl}
\zeta^{(2)}_\sigma =\frac{1}{2} \left(\zeta^{(1)}_\sigma \right)^2=
\frac{1}{2} \left(\zeta^{(1)}_\sigma \right)_I^2\, ,
\ee
where we have emphasized that also $\zeta^{(2)}_\sigma$is a conserved quantity
whose value is determined by the curvaton fluctuations during inflation. 
Plugging Eq.~(\ref{zeta2infl}) into Eq.~(\ref{zeta2decay}) the 
curvature perturbation during the standard radiation or matter dominated 
eras turns out to be
\be
\label{zetafinal}
\zeta^{(2)}=f_D \left( \frac{3}{2}-f_D^2 \right) 
\left( \zeta^{(1)}_\sigma \right)^2\, .
\ee


\subsection{Non-Gaussianity of the curvaton perturbations}
\label{non-gauss}
Let us now now 
focus on the calculation of the non-linear parameter $f_{\rm{NL}}$ which is 
usually adopted to characterized the level of non-Gaussianity of the Bardeen 
potential \cite{ks}. In order to compute $f_{\rm{NL}}$ 
in the curvaton scenario, we   
switch from the spatially flat gauge $\psi=0$ to the longitudinal or 
Poisson gauge \cite{poisson}. Such a procedure is 
possible since the curvature perturbations 
$\zeta^{(2)}_i$ are gauge-invariant quantities. 
In particular this is evident form the expression 
found in Eq.~(\ref{zetafinal}). 
During the matter dominated era from Eq.~(\ref{qqq}) it turns out that 
\cite{BMR} 
\bea
\label{zetam}
\zeta^{(2)}&=&-\psi^{(2)}+\frac{1}{3} \frac{\delta^{(2)}\rho}{\rho}+
\frac{5}{9} \left( \frac{\delta^{(1)}\rho}{\rho} \right)^2 \nonumber\\
&=&-\psi^{(2)}+\frac{1}{3} \frac{\delta^{(2)}\rho}{\rho}+\frac{20}{9} 
\left( \psi^{(1)} \right)^2\, , 
\eea
where in the last step we have used that on large scales 
$\delta^{(1)}\rho /\rho=-2 \psi^{(1)}$ in the Poisson gauge 
\cite{BMR}.
Eq.~(\ref{zetam}) combined with Eq.~(\ref{zetafinal}), 
which gives the constant value on large scales
of the curvature perturbation $\zeta^{(2)}$ during the matter dominated era, 
yields
\be
\psi^{(2)}-\frac{1}{3} \frac{\delta^{(2)} \rho}{\rho}=\frac{1}{9}
\left[ 20-\frac{75}{2 f_D}+25 f_D \right] \left( \psi^{(1)} \right)^2\, ,
\ee
where we have used $f_D \zeta^{(1)}_\sigma=-\frac{5}{3} \psi^{(1)}$ from 
Eq.~(\ref{atcurvdecay}) and the usual linear relation between the curvature 
perturbation and the Bardeen potential $\zeta^{(1)}=-\frac{5}{3} \psi^{(1)}$ 
during the matter dominated era. Since on large scales
(from the second-order  $(0-0)$- and $(i-j)$-components of Einstein equations,
see Eqs. (A.39) and (A.42-43)  in \cite{noi}) 
the following relations hold during the matter-dominated
phase

\begin{eqnarray}
\phi^{(2)}&=&-\frac{1}{2}\frac{\delta\rho^{(2)}}{\rho}+
4\left(\psi^{(1)}\right)^2\, ,\nonumber\\
\psi^{(2)}-\phi^{(2)}&=&-\frac{2}{3}\left(\psi^{(1)}\right)^2+
\frac{10}{3}\nabla^{-2}\left(\psi^{(1)}\nabla^2\psi^{(1)}\right)\nonumber\\
&-&10\,\nabla^{-2}\left(\partial^i\partial_j\left(\psi^{(1)}\partial_i
\partial^j
\psi^{(1)}\right)\right)\, ,
\end{eqnarray}
we conclude that

\begin{eqnarray}
\label{final}
\phi^{(2)}&=&\left[\frac{10}{3}+\frac{5}{3}f_D-\frac{5}{2 f_D}\right]
\left(\psi^{(1)}\right)^2 \nonumber \\
&-&2 \nabla^{-2}\left(\psi^{(1)}\nabla^2\psi^{(1)}\right)
+6\,\nabla^{-2}\left(\partial^i\partial_j\left(\psi^{(1)}\partial_i
\partial^j
\psi^{(1)}\right)\right)\, . \nonumber \\
&&
\end{eqnarray}
The 
total curvature perturbation will then have a
non-Gaussian
$(\chi^2)$-component. The lapse function 
$\phi=\phi^{(1)}+\frac{1}{2}\phi^{(2)}$ can be expressed in momentum space
as 
\begin{eqnarray}
\phi({\bf k}) &=& \phi^{(1)}({\bf k})+ 
\frac{1}{(2\pi)^3}
\int\, d^3 k_1\,d^3 k_2\, \delta^{(3)}\left({\bf k}_1+{\bf k}_2-{\bf k}\right)
\nonumber\\
&\times&
f^\phi_{\rm NL}\left({\bf k}_1,{\bf k}_2\right)
\phi^{(1)}({\bf k}_1)\phi^{(1)}({\bf k}_2)\, ,
\label{gauss}
\end{eqnarray}
where we have defined 
an effective ``momentum-dependent''
non-linearity parameter $f^\phi_{\rm NL}$.
Here the linear lapse function 
$\phi^{(1)}=\psi^{(1)}$ is a Gaussian random field. 
The gravitational potential bispectrum
reads

\begin{eqnarray}
&&\langle \phi({\bf k}_1) \phi({\bf k}_2) \phi({\bf k}_3)
\rangle=(2\pi)^3\,\delta^{(3)}\left({\bf k}_1+{\bf k}_2+{\bf k}_3\right)
\nonumber\\
&&\times \,\left[2\,f^\phi_{\rm NL}
\left({\bf k}_1,{\bf k}_2\right)\,
{\cal P}_\phi(k_1){\cal P}_\phi(k_2)+{\rm cyclic}\right]\, ,
\end{eqnarray}
where ${\cal P}_\phi(k)$ is the power-spectrum of the gravitational
potential. From Eq.~(\ref{final}) we read the the non-linearity parameter

\begin{eqnarray}
\label{ee}
f^\phi_{\rm NL}=\left[\frac{7}{6}+\frac{5}{6}r-\frac{5}{4 r}\right]+ 
g({\bf k}_1,{\bf k}_2) \, ,
\end{eqnarray}
where 
\be
g({\bf k}_1,{\bf k}_2) = \frac{{\bf k}_1 \cdot {\bf k}_2}
{k^2}\left(1+ 3\frac{{\bf k}_1\cdot {\bf k}_2}{k^2} \right)
\, ,
\ee
with ${\bf k} ={\bf k}_1 +  {\bf k}_2$ and 
we have replaced $f_D$ with $r$ to go beyond the sudden approximation. 
Notice that in the final bispectrum expression, the diverging terms arising 
from the infrared behaviour of $f^\phi_{\rm NL}({\bf k}_1,{\bf k}_2)$ 
are automatically regularized once the monopole term is subtracted from the 
definition of $\phi$ (by requiring that $\langle \phi \rangle$=0). 

As far as the momentum-independent part is concerned we note that 
in the limit $r\ll 1$ we obtain $f^\phi_{\rm NL}=
-\frac{5}{4 r}$ which reproduces the estimate provided in \cite{LW,LUW},
while, in the limit $r\simeq 1$, we obtain $f^\phi_{\rm NL}=\frac{3}{4}$
\footnote{Notice that the formula (36) in  \cite{LUW}
for the estimate of the
non-linear parameter contains a sign misprint and 
should read $f^\phi_{\rm NL}\simeq -\frac{5}{4 r}$, giving 
$f^\phi_{\rm NL}\simeq -\frac{5}{4}$ for $r\simeq 1$.}
for $r\simeq 1$. 
These values have to be compared to the value $f^\phi_{\rm NL}=-\frac{1}{2}$
\cite{BMR} 
obtained for perturbations whose wavelengths re-enter the horizon during
the matter-dominated phase for the standard scenario in which
curvature perturbations are induced by fluctuations of the inflaton field.
We conclude that if $r \ll 1$ the non-Gaussianity in the curvaton scenario
is larger than the one predicted in the standard scenario.
Finally we point out that additional non-Gaussianity will be generated 
after horizon-crossing, due to known Newtonian and relativistic 
second-order contributions which are  
relevant  on sub-horizon scales, such as the Rees-Sciama effect~\cite{rs}, 
whose detailed analysis has been given  in Refs.~\cite{pyne}. 
It is important to consider also these effects 
when making a comparison with observations.  

\section{A comment on the evolution of the 
curvature perturbation during the reheating phase in the standard scenario}
\label{comment}
Note that Eq.~(\ref{main}) 
is indeed valid in the general framework of an
oscillating scalar field and a radiation fluid, the curvaton 
scenario being only a particular case. Thus in this section we shall 
indicate the generic scalar field by $\varphi$ instead of $\sigma$. 
From Eq.~(\ref{main}) it is possible to derive the following 
equation of motion on large scales for the  second-order  
curvature perturbation $\zeta^{(2)}$
\be
\zeta^{(2)'}=f'\left( \zeta^{(2)}_\varphi-\zeta^{(2)}_\gamma\right)
+f' (1-3f^2) \left( \zeta^{(1)}_\varphi-\zeta^{(1)}_\gamma\right)^2\, ,
\ee
where we have used the fact that, in the approximation of sudden decay of 
the scalar field $\varphi$, the 
individual curvature perturbations at first and second-order 
are separately conserved.
Using Eqs.~(\ref{zetasum}) and (\ref{main}) it is possible to rewrite 
$\zeta^{(2)'}$ in terms only of $\zeta^{(2)}$, $\zeta^{(2)}_\varphi$ and 
$\zeta^{(1)}$, $\zeta^{(1)}_\varphi$ as 
\be
\label{zeta2'}
\zeta^{(2)'}=-\H f \left( \zeta^{(2)}-\zeta^{(2)}_\varphi \right)
+\H f(1+2f)\left( \zeta^{(1)}-\zeta^{(1)}_\varphi \right)^2\, .
\ee
Here we want to make a simple but important observation.
Besides the curvaton scenario, the most  interesting case where there is 
an oscillating scalar field and a  radiation fluid is just the 
phase of reheating following a period of inflation in the standard
scenario for the generation of cosmological perturbations on large
scales.  
In such a situation the oscillating scalar field is just the inflaton 
field $\varphi$ whose fluctuations induce curvature perturbations.  
Therefore it is possible to see in a straightforward way that 
during the reheating phase, when the inflaton field finally decays into 
radiation, 
a solution of Eq.~(\ref{zeta2'}) is the one corresponding to a total 
curvature perturbation which is indeed fixed by the inflaton 
curvature perturbation during inflation $\zeta^{(1)}=\zeta^{(1)}_\varphi=
\zeta^{(1)}_{\varphi I}$,
$\zeta^{(2)}=\zeta^{(2)}_\varphi=\zeta^{(2)}_{\varphi I}$.


\section{Conclusions}
In this paper we have determined the evolution on large scales of the
second-order curvature perturbation within the curvaton scenario where
two fluids are present, the curvaton and radiation. We have computed the
the non-linear parameter $f_{\rm NL}$ measuring the
level of non-Gaussianity in the primordial cosmological perturbations
and provide its exact expression as a function of the parameter $r$. 
Our findings are particularly interesting if one wishes to  extract from
a  positive future detection of non-linearity
in the CMB anisotropy pattern a way to 
descriminate
among the mechanisms
by which  cosmological adiabatic perturbations are generated.
It would be interesting to extend our results to those models
which can accomodate for a primordial value of $f_{\rm NL}$ 
larger than unity. This is the case, for instance, of a large
class of multi-field inflation models which leads to either 
non-Gaussian isocurvature perturbations \cite{lm} or cross-correlated  
non-Gaussian adiabatic and isocurvature modes \cite{bartolong}
and the so-called
``inhomogeneous reheating'' mechanism where the curvature
perturbations are generated by spatial variations of the
inflaton decay rate \cite{gamma}.

\end{document}